\RequirePackage{fix-cm}
\documentclass[final,natbib]{svjour3}                     
\smartqed  
\usepackage{mathptmx} 
\usepackage{amsmath, amssymb, amsfonts}
\usepackage{latexsym}
\usepackage{tabulary}
\usepackage{array,longtable}
\usepackage{url}
\usepackage{doi}

\usepackage{graphicx}

\usepackage{epsfig}
\usepackage{epstopdf}
\journalname{}
\begin{document}

\title{Achieving ``space of physics journals'': topological structure and the Journal Impact Factor 
}


\author{Yurij L. Katchanov \and
        Yulia V. Markova
}

\institute{Y. L. Katchanov \at
            National Research University Higher School of Economics (HSE) \\
			20 Myasnitskaya Ulitsa, Moscow 101000, Russia \\
            \email{yurij.katchanov@gmail.com}           
           \and
            Y. V. Markova \at
            American Association for the Advancement of Science \\
			1200 New York Ave NW, 20005 Washington, DC, USA \\
            \email{yulia.markova@gmail.com}           
}

\date{no date}

\maketitle

\begin{abstract}
The empirical distribution function of citations to journal articles (EDF for short) can become the fundamental tool for analyzing the scientific journals. Endeavors at making bibliometric analysis independent of the intuition of average citation levels have led us to the study of qualitative properties of physics journals in the functional space of EDFs. We show that the structure of this space establishes the connections and relationships that determine the essential features of physics journals. The research provides an analysis of 240 physics journals indexed in Journal Citation Reports 2015. The relevance of EDFs clustering is discussed. Our findings reveal four-cluster space of physics journals. The space brings to light the essential distinctions between physics journals and shows different level of influence of scientific publishers  belonging to different types (professional physics societies, transnational and local publishers). The study of EDFs grouped by publishers reveals two binary oppositions that structure relations between them: ``global\,--- local'' publishers and ``high cited\,--- low cited'' publishers. 

\end{abstract}
\maketitle

\section*{Introduction}
In its present form, bibliometrics still cannot take the place of in-depth science of science (see, e.g., \cite{Boerner2012, cronin_beyond_2014} for details). Bibliometrics does not offer a ``clear window'' into science, but rather something more akin to a stained-glass window, which allows one to gain some impressions of the world, but at the same time imposes its own patterns and colours on that world. In this regard, the scare quotes around the term ``space of physics journals'' (hereafter denoted as SPJ) in the title of this paper are intentional~--- they draw attention to the fact that we must explicitly define what we mean by the term.

The ``bibliometric expansion'' of recent years~--- the rapid influx of bibliometric indicators into science policy and management~--- has led to many scientometric concepts becoming integrated into general scientific language. The prevailing view is that scientometrics enables us to discuss more clearly and precisely issues related to the evaluation of universities, laboratories, individual scientists and scientific journals. However, as a practical knowledge \cite{Kinouchi_2014, Radicchi2015, Vitanov2016, Mingers2017323, ESTS115} scientometrics serves only as a subsidiary element of the system of scientific governance \cite{Beer2016}. The cognitive value of the vast majority of scientometric research on scientific journals is limited to various types of indicators (we do not go into this question here, and refer the reader to, e.g., \cite{Glaenzel2013, Mingers, Kosten2016,  Leydesdorff2016, Rijcke_2016, Bornmann2017}), with the most well-respected being the Journal Impact Factor \cite{10.1001/jama.295.1.90} (JIF for short). It is provided by the Journal Citations Reports (Clarivate Analytics), and is a quantitative measure for ranking and evaluating scientific journals. It consists of ``the average number of citations received per paper published in that journal during the specific period of two preceding years'' \cite{reuters2012thomson}. JIF is ``a ratio between citations and recent citable items published'' \cite{reuters2012thomson}. The JIF calculation is well presented in the bibliometric literature, therefore we shall not pursue it here.

JIF has become practically the standard measure for evaluating and ranking scientific journals (see \cite{Moed2005, Mingers}, and \cite{todeschini_2016}, among others). The receipt of a high JIF is an object of fierce competition between scientific publishers; this indicator is actively used in scientific policy and management. Unsurprisingly, JIF has been subject to close scrutiny from professional and citizen bibliometricians, as well as researchers, publishers, policymakers and research managers. Since an analysis of the literature on JIF is not one of the aims of our work, we refer the reader to several excellent reviews which consider this subject in depth \cite{glanzel_journal_2002, bar-ilan_2008, Waltman2016365, 10.3389/frma.2017.00004}. Despite being the subject of ongoing criticism (see, inter alia, \cite{Finardi_2013, Bornmann1094, Chorus2016, 10.1371/journal.pone.0151414, Shanahan_2016, Chua2017, Gasparyan2017}), JIF continues to be widely used (e.g., \cite{Kiesslich_2016, Tahamtan2016, Pudovkin2017, Mila_2017, Waltman_arXiv2017, Zhang2017}). If we accept the principle of sufficient reason, then we must postulate that JIF has a reason for being (cf. \cite{Paulus_2015, Beer2016}). In this paper, the point is not so much JIF of any given journal itself, but rather understanding how JIF works.

Bourdieu's idea that social sciences ``present itself as a \emph{social topology}'' \cite{Bourdieu1985} becomes the basis of many attractive theoretical models. As scientometrics is one of social sciences, it seems promising to search for those scientometric regularities that can be described in the language of topology. 

From this ``topological'' point of view, it is necessary to realize that JIF of a scientific journal has practical significance in citation field only in reference to JIFs of other journals. Since the issue we are seeking to address is concerned not with individual physics journals, but rather with the sample \(S\) of physics journals and the interrelations between them, it should come as no surprise that the resultant working model is a \emph{topological structure} \(\mathcal{T}\). It is in our case a collection of families of clusters (which has properties of neighbourhood system for SPJ \((S, \mathcal{T})\)). Roughly speaking, elements of the topological structure \(\mathcal{T}\) are unions of open balls in a suitable metric space of empirical distribution functions of citations. The concept of ``topological structure'' (see the formal definition in \cite{elements}) is a means of studying the qualitative properties of the totality of physics journals. The \emph{topological space} of physics journals is a mathematical construct that relates the journals within the entire bibliometric data set to each other, and thereby produces a single scientometric entity \((S, \mathcal{T})\). It is obtained by taking a sample \(S\) of physics journals and equipping \(S\) with \(\mathcal{T}\), by defining relations between these journals.

Since present paper is concerned with the results of bibliometric measurements, and results may be regarded as a random variable, it is not surprising that the mathematical tools required to describe these variables at a fundamental level should be the empirical (cumulative) distribution functions of citations to journal articles (cf. \cite{Lariviere062109, Blanford2016}). For the bibliometric concept of an indicator, we must substitute the concept of a state of a journal. The scientometric state of a physics journal is characterized by an empirical (cumulative) distribution function, which contains all the statistical information it is possible to obtain about the citations to the journal.

For convenience of expression, we denote by \(\hat{F}_{\xi}\) the empirical distribution function of citations (or, in abbreviated form, EDF), corresponding to the journal \(\xi\). We lay great stress on \(\hat{F}_{\xi}\) because this function becomes infinitely close to the original distribution function \(F_{\xi}\) of the observed random variable (for a full treatment of this subject, see \cite{Borovkov:1998}). Our key intuition is to think of \(\hat{F}_{\xi}\) as a state of the journal \(\xi\). To give a scientometric meaning to the EDF, we assume that a physics journal \(\xi\) is conceptualized as random variable \(\xi(\omega)\), and \(\hat{F}_{\xi}\) represent the state of the journal \(\xi\). This interpretation will serve as a guideline throughout this paper.

Our tasks in this paper are threefold:
\begin{enumerate}
\item To study EDFs as a bibliometric characterization of physics journals problem.
\item To adopt distances in the normalized Kolmogorov metric to introduce a topological structure on the sample of physics journals for solving the clustering problem.
\item To provide an analysis of the sample of 240 physics journals, in the context of the relationships and connections that exist between EDFs of these journals.
\end{enumerate}
The main contributions of this paper are the following:
\begin{itemize}
\item We describe the sample of physics journals. We approach this by constructing what we call the ``space of physics journals''.
\item We provide an analysis of cluster structure of SPJ using such journals' properties as size, country, publisher, and so on.
\end{itemize}

\section*{Data and methods}
The data on journal citations and indicators was extracted from the Web of Science Core Collection (WoS CC for short)~--- scientific citation indexing service, and the Journal Citation Reports~--- an analytic tool yields bibliometric information about academic journals. Both are provided by Clarivate Analytics. The first step was to select all journals included in the ``physics'' category in the Journal Citation Reports, according to the category schema ``Essential Science Indicators'', which resulted in \(298\) items. Journals which had published less than \(100\) papers in two years were excluded from the selection, resulting in a narrowed sample of \(240\) journals (see Tab.~\ref{tab1}). It is worth noting that six of the ten journals with the highest JIF were excluded from the sample, since these were review journals publishing a small number of papers per year. Taken together, the excluded journals publish on average 55.8 citable items in two years. The next step was to extract information from the WoS CC on citations made in 2015 for all papers published in the journal sample during 2013 and 2014. Only ``article'' and ``review'' data types were included in the sample. We opted for a two year publication window, a one year citation window, and ``article'' and ``review'' document types, so that our corpus would be as close as possible to that which forms the basis for the calculation of JIF in the Journal Citation Reports.

The dataset was downloaded from WoS CC in August 2016. In addition to citation data, we used journal indicators provided by the Journal Citation Reports for the year 2015. Data on journals' publishers and countries were extracted from the ``Scopus title list''.

For statistical data processing, we used the R programming language and IBM SPSS Statistics\,22.

\section*{Theory}
In this paper, we treat the word ``theory'' approximately as the ``formalism'', that is, a collection of abstract analytic tools for working formally with bibliometric quantities, deriving formulas, and interpreting them. General and systematic formalism is almost as useful as exact formulas.

\subsection*{Space of EDFs}

We postulate that EDFs form the basis of the empirical representation of physics journals. This abstract principle establishes a link between mathematical and scientometric objects:
\begin{enumerate}
\item A state of a physics journal \(\xi\), defined by its measurement~--- the resulting set \(X_{n} = (x_{1},\dots , x_{n})\) of \(n\) observations of the publication\,--\,citation process, is represented by EDF \(\hat{F}_{\xi n}(x)\) of \(\xi\) in \(n\) trials, where the index \(n\) will be left out.
\item The mathematical image of a totality of physics journals is a convenient linear functional space \(D_{b}\).
\end{enumerate}
It is instructive here to compare bibliometric indicators with EDF:
\begin{itemize}
\item bibliometric indicators are various numerical sample characteristics,
\item the EDF \(\hat{F}_{\xi}(x)\) completely specifies the probability distribution of the random variable \(\xi(\omega)\).
\end{itemize} 
The realization that the physics journals are not represented by real numbers (bibliometric indicators referring to the journals) but by functions (EDFs), is one of our major findings. The EDF \(\hat{F}_{\xi}\) contains the essential information about the journal \(\xi\). 

The above idea can be expressed more precisely by means of the notion of a metric space: the EDF \(\hat{F}_{\xi}\) is a point of the subspace \((D_{b}, \rho)\) of the Skorokhod space, comprising all bounded, nondecreasing functions that are right-continuous and have left-hand limits, equipped with the normalized uniform (or Kolmogorov) metric:
\begin{equation}
\label{dist}
\rho (\xi, \eta) = \sqrt{\frac{nm}{n + m}} \sup_{x}\left\lbrace\bigl| \hat{F}_{\xi}(x) - \hat{F}_{\eta}(x)\bigr|\right\rbrace ,
\end{equation}
where \(\hat{F}_{\xi}\) and \(\hat{F}_{\eta}\), corresponding to the journals \(\xi\) and \(\eta\), are EDFs based on the samples \(X_{i}\) and \(Y_{j}\) of sizes \(n\) and \(m\), respectively. We have glossed over some details here, and refer the reader to \cite{Jacod2003, RS2013}. To help the intuitive interpretation of what follows, we shall treat the space \(D_{b}\) as a phase space. If one adopts this point of view, the state of a journal on \(D_{b}\) is defined by a point \(\hat{F}_{\xi}\). Thus, the EDF \(\hat{F}_{\xi}\) is a point of space \(D_{b}\), and at the same time possesses an inner structure.

Let us indicate with \(S\) the sample of physics journals such that \(S \subset D_{b}\). The formal definition of the \emph{deviation} from a journal \(\xi\) to \(S\) is the following: 
\begin{equation}
\label{pj}
d(\xi) = \inf\left\lbrace \rho (\xi, \eta) \colon \hat{F}_{\eta} \in S \setminus {\hat{F}_{\xi}} \right\rbrace,
\end{equation}
where \(S \setminus {\hat{F}_{\xi}}\) denote the difference between \(S\) and \(\hat{F}_{\xi}\). We can see more clearly the geometrical significance of the above definition. The formula \eqref{pj} means that \(d(\xi)\) is the distance from \(\hat{F}_{\xi}\) to the set \(S \setminus {\hat{F}_{\xi}}\) corresponding to the sample \(S\). Incidentally, we may notice that \(|d(\xi) - d(\eta)| \leqslant \rho (\xi, \eta)\).

\subsection*{Partial order of Bishop\,--\,Phelps}
Let us now be more technical. One knows that \(D_{b}\) is a Banach space, and we use \(D^{\star}_{b}\) and \(\langle\cdot ,\cdot\rangle\) to denote its topological dual and the duality pairing, respectively. This yields that the dual \(D^{\star}_{b}\) of \(D_{b}\) can be considered as the space of utilities \(U(\cdot)\) associating with any journal \(\xi\) (i.e., \(\hat{F}_{\xi}\in D_{b}\)) its value \(\langle \hat{F}_{\xi}^{\star}, \hat{F}_{\xi}\rangle = U(\hat{F}_{\xi})\).

Yet \(D_{b}\) may not have any order structure. For most scientometric problems of interest it is possible to introduce a rather stronger concept incorporating a partial order. Choose any \(\varepsilon > 0\); then, the partial order of Bishop\,--\,Phelps on \(D_{b}\times\mathbb{R}_{+}\) can be defined as follows (cf. \cite{Phelps_1993}):
\begin{equation}
\label{bp}
\left[\hat{F}_{\xi},\hat{F}_{\xi}^{\star}\right] \precsim \left[\hat{F}_{\eta},\hat{F}_{\eta}^{\star}\right] \Leftrightarrow 
 U \bigl(\hat{F}_{\xi}\bigr) + \varepsilon \rho (\xi, \eta) \leq  U \bigl(\hat{F}_{\eta}\bigr) .
\end{equation}
We get the formula \eqref{bp} out of our general assumptions without any arbitrariness. Actually, the utility function is uniquely determined by the geometry of the space \(D_{b}\), since for \(\rho(\cdot, \cdot)\) we can construct the functional \(U(\cdot)\) with the required properties. By construction, the expression
\(\langle \hat{F}_{\eta}^{\star}, \hat{F}_{\eta}\rangle - \langle \hat{F}_{\xi}^{\star}, \hat{F}_{\xi}\rangle = U(\hat{F}_{\eta}) - U(\hat{F}_{\xi})\)
is a ``gain'' of the journal \(\eta\) \emph{relative} to the journal \(\xi\).

\subsection*{Choice of utility function}
The bibliometric agenda has for a long time, to a significant extent, been dominated by concerns around JIF. Moreover, JIF is one of the major stakes in the ``citations game'' being played by physics journals. It follows that there must be good reasons for this. This state of affairs allows us to take the mean value of citations \(\mathsf{E}_{\xi}\) to a journal \(\xi\) as the utility function of \(\hat{F}_{\xi}\) (cf. \cite{Seiler2014904}). Substituting \(\mathsf{E}_{\xi}\) for \(U(\hat{F}_{\xi})\) in \eqref{bp}, we get
\begin{equation}
\label{po}
 \hat{F}_{\xi} \precsim \hat{F}_{\eta} \Leftrightarrow \mathsf{E}_{\xi} + \varepsilon \rho (\xi, \eta) \leq  \mathsf{E}_{\eta} .
\end{equation}
The meaning of this expression is quite understandable (cf. \cite{Bouyssou2014}).

\subsection*{Finite topological space of EDFs}
Though the topological structure \(\mathcal{T}\) on \(S\) involved is quite laborious, we can obtain the desired results step by step (the following is slightly adapted from \cite{Barmak2011_1, may2008}).
\begin{enumerate}
\item{Empirically, we divide \((S, \rho)\) into maximal (with respect to inclusion) indiscrete subspaces or just clusters \(C_{\alpha}\) of EDFs:
\begin{equation}
\label{cl1}
(S\subset D_{b})\left(\forall \alpha \ne \beta\colon C_{\alpha}\bigcap C_{\beta} = \emptyset\right)\colon S = \bigcup_{\alpha} C_{\alpha}.
\end{equation}
It is commonly held that an indiscrete cluster \(C_{\alpha}\) equipped with so-called trivial topological structure, which consist only of \(C_{\alpha}\) and \(\emptyset\).}
\item{Let us call all the EDFs from \(S\) that fall into any cluster \(C_{\alpha}\) equivalent and then identified to one element \(c_{\alpha}\). The index \(\alpha\) simply tells us which cluster we are talking about. However, the quotient set \(S/{\thicksim} \ = \bigcup_{\alpha} c_{\alpha}\) with respect to the given equivalence relation \(\thicksim\) is not a subspace of \(D_{b}\), and is not equipped with the metric \(\rho(\cdot ,\cdot)\). Any continuous map \(\phi\colon (S, \mathcal{T}) \rightarrow (S^{\prime}, \mathcal{T}^{\prime})\) respect the clusters \(C_{\alpha}\). The topological structure \(\tau\) of the quotient set \(S/{\thicksim}\) is generated by the base consisting of the sets of the form \(\mathsf{C}_{S}^{+}(c_{\alpha}) = \lbrace s\in S/{\thicksim}\colon c_{\alpha} \precsim s \rbrace\), where the symbol \(\precsim\) denotes the partial order~\eqref{po}. \((S/{\thicksim}, \tau)\) is a smallest neighborhood space, i.e. each element \(s\in S/{\thicksim}\) has the smallest (with respect to inclusion) neighborhood \(\mathsf{C}_{S}^{+}(c_{\alpha})\). It has been known that \((S, \mathcal{T})\) is determined up to homeomorphism by \((S/{\thicksim}, \tau)\). Furthermore, \((S, \mathcal{T})\) and \((S^{\prime}, \mathcal{T}^{\prime})\) are homeomorphic iff there exists a monotone bijective quotient map \(\varphi\colon (S/{\thicksim}, \tau) \rightarrow (S^{\prime}/{\thicksim^{\prime}}, \tau^{\prime})\). This yield that SPJ is, up to homeomorphism, a finite partial order set of indiscrete clusters of physics journals. For an enlightening discussion of these questions the reader is referred to \cite{Barmak2011_1}.}
\item{It is possible to indicate that \((S, \rho)\) is metrizable because it is discrete. At that, the minimal neighborhood base of \((S, \rho)\) at a point \(\hat{F}_{\xi} \in S\) is the open ball \(B\bigl(\xi, d(\xi)\bigr) = \bigl\lbrace S \ni \hat{F}_{\eta} \colon \rho (\xi, \eta) < d(\xi) \bigr\rbrace \). In accordance with standard practice we say that a family \(\bigl(B_{i}(\cdot, d(\cdot))\bigr)_{i\in I}\) forms the minimal base for the open sets of \((S, \rho)\), i.e. every open set of \((S, \rho)\) is the union of the subfamily of the family \(\bigl(B_{i}(\cdot, d(\cdot))\bigr)\). Hence, it is tempting in some contexts to regard \(d(\xi)\) as related to the minimal base \(\bigl(B_{i}(\cdot, d(\cdot))\bigr)_{i\in I}\).}
\item{The Hausdorff distance between clusters \(A\) and \(B\) is defined by letting
\begin{equation}
\label{hausdorff}
\rho_{\mathrm{H}}(A, B) = 
\max\left\lbrace \sup_{\alpha\in A} \: \inf_{\beta\in B} \rho \bigl(\alpha, \beta\bigr), \: \sup_{\beta\in B}\: \inf_{\alpha\in A} \rho \bigl(\alpha, \beta\bigr)\right\rbrace .
\end{equation} }
\end{enumerate}

\section*{Results and discussion}
The formulation of the topological structure as a (relatively) autonomous scientometric entity (i.e. based only on itself) provides an imperative to uncover JIF through topological concepts. We first begin our discussion with a formal but consequential result: the deviation \(d(\xi)\) is highly correlated with \(\mathrm{JIF}(\xi)\) (Pearson's \(R = 0{.}879\), \(p = 0{.}000\)). The empirical relationship between quantities \(d(\xi)\) and \(\mathrm{JIF}(\xi)\) bear a linear character and can perhaps be expressed by the following regression equation:
\begin{equation}
\label{regr}
\mathrm{JIF}(\xi) = 9{.}758\cdot d(\xi) - 0{.}824 \quad (R^{2} = 0{.}772, \: p = 0{.}000).
\end{equation}
The formula \eqref{regr} clearly shows that \(\mathrm{JIF}(\xi)\) is proportional to the deviation \(d(\xi)\). We already know that \(d(\xi)\) characterizes the structure of citations of a journal \(\xi\) relative to the sample \(S\). It must be kept in mind when we come to interpret Eq.~\eqref{regr}. This implies that our approach provides a framework for thinking of the ranking of physics journals according to their JIF as a result of the topological structure.

\begin{figure}[ht!]
\centering
  \includegraphics[width=\linewidth]{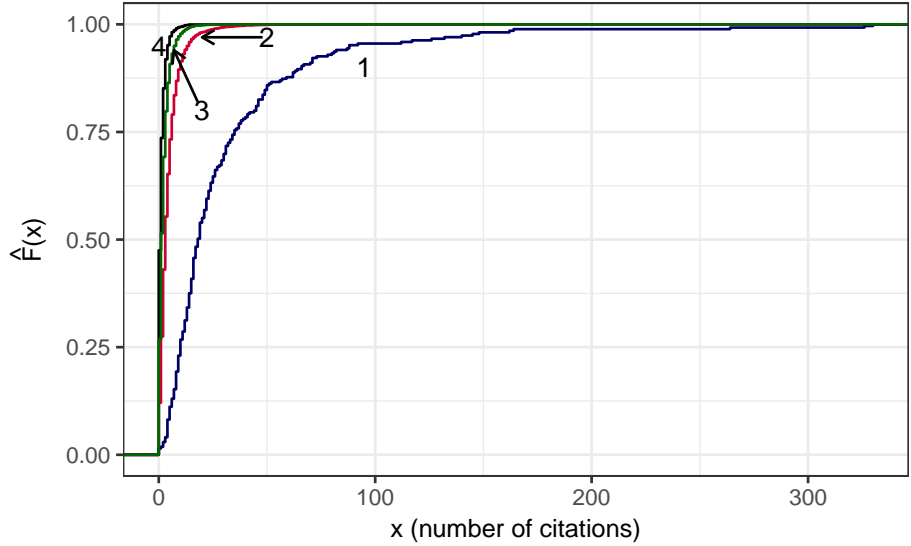}
\caption {The EDFs for the centers of the clusters: 1~Nature Photonics (cluster\,1), 2~Journal of Physical Chemistry C (cluster\,2), 3~Physical Review E (cluster\,3); 4~IEEE Transactions on Magnetics (cluster\,4)}
\label{fig:1}      
\end{figure}

The cluster analysis of the sample \(S\) is done in two steps (cf., e.g., \cite{ASI23370, Varin_2016, vanEck2017}).
\begin{enumerate}
\item{At the first stage, we evaluate the number of clusters.
\begin{itemize}
\item{To this end, every EDF \(\hat{F}_{\xi} \in S\) is paired with a set \(s[\xi]\) of all nearest EDFs to \(\hat{F}_{\xi}\) such that the following inequality is satisfied:
\begin{equation}
\label{cr}
\bigl( \hat{F}_{\xi}, \hat{F}_{\eta} \in S \bigr) \colon \inf_{\lambda}\left\lbrace \lambda\colon \mathrm{P}\left(\rho \bigl(\xi, \eta\bigr) > \lambda \right) < \lambda \right\rbrace \leq \Lambda .
\end{equation}
Here, \(\mathrm{P}(\cdot)\) denotes a probability, and \(\Lambda\) is a positive constant. We can obtain \(\mathrm{P}(\cdot)\) in \eqref{cr} by applying Smirnov's theorem \cite[p.~384]{Borovkov:1998}. 
}
\item{Fix \(\Lambda = 0{.}75\) and take the greatest element \(\hat{F}_{\max}\) of \(S\). Here, it is important to remind that \(S\) is partially ordered by \(\precsim\) \eqref{po}. By straightforward calculations we find that \(\hat{F}_{\max} = \hat{F}_{149}\). It is the center of cluster 1 (or a cluster 1 prototype). All EDFs that fall into \(s[149]\) are eliminated from \(S\). The greatest element of remaining EDFs \(\hat{F}_{109}\) is the center of cluster 2. At this stage, all elements of \(s[109]\) are eliminated from \(S\). Then this procedure is repeated two times. The centers of clusters 3 and 4 are \(\hat{F}_{184}\) and \(\hat{F}_{73}\), respectively. Fig.~\ref{fig:1} demonstrates the centers of clusters. Using above procedure, we determine that the sample \(S\) can be divided into four clusters.}
\end{itemize}
}
\item{The four clusters \(\bigl(\mathrm{Cl}_{j}\bigr)_{j = 1}^{j = 4}\) of physics journals were formed using the normalized Kolmogorov distance matrix \(\bigl( \rho (\xi, \eta)\bigr)_{\xi, \eta =1}^{\xi , \eta = 240}\) and the ``ward.D2'' agglomeration method \cite{Murtagh2014} (the function hclust of R package stats).}
\end{enumerate}
Let us begin with the simple observation that we obtain the clusters by applying topological concepts within the statistical description of SPJ. Also, we can characterize topological aspects of clusters. Consider a cluster \(\mathrm{Cl} \in \bigl(\mathrm{Cl}_{j}\bigr)_{j \in J}\). Define the following
\begin{align}
\varrho_{\mathrm{cover}} (\mathrm{Cl}) &= \max_{\xi \in S} \min_{\eta \in \mathrm{Cl}} \rho (\eta, \xi) , \label{R1} \\
(\xi \ne \eta)\colon \rho_{\min} (\mathrm{Cl}) &= \min_{\xi, \eta \in (\mathrm{Cl})} \rho (\xi, \eta) , \label{R2} \\
\varrho_{\mathrm{C}} (\mathrm{Cl}) &= \min_{\xi \in S} \max_{\eta \in \mathrm{Cl}} \rho (\eta, \xi) . \label{R3}
\end{align}
The quantities introduced above are called covering radius \eqref{R1}, minimum distance \eqref{R2}, and Chebyshev radius \eqref{R3} of \(\mathrm{Cl}\), respectively. Note that the Chebyshev radius of \(\mathrm{Cl}\) is the minimal radius of an open ball containing \(\mathrm{Cl}\).

The brief Tab.~\ref{tab:0} illustrates some aspects of the clusters of physics journals. It can be seen from the data in Tab.~\ref{tab:0} that \(\varrho_{\mathrm{cover}}(\mathrm{Cl})\) and \(\rho_{\min}(\mathrm{Cl})\) decrease monotonically from cluster 1 to cluster 4. It gives evidence that EDFs of cluster 1 are less homogeneous than EDFs of cluster 2 and EDFs of cluster 2 are less homogeneous than EDFs of cluster 3 and so on. On the other hand, \(\varrho_{\mathrm{C}}(\mathrm{Cl})\) shows more complex behavior. Nevertheless, the Chebyshev radiuses may indicate the quality of our clustering. In fact, the sum of the Chebyshev radiuses of four clusters is equal \(29{.}805\), whereas radius of \(S\) is \(33{.}056\). In other words, the sample covers all four clusters.

\begin{table*}[!htb]
\centering
\caption{Some characteristics of clusters}
\label{tab:0}
\begin{tabular}{lllllll}
\hline
No. & Center of cluster & \(\varrho_{\mathrm{cover}}\) & \(\rho_{\min}\)	& \(\varrho_{\mathrm{C}}\)	& \% of  	& \% of citations\\
	&					&							 &					& 							& journals  & in 2015 \\
\hline
1 & Nature Photonics    & \(2{.}209\) 				& \(0{.}670\) 		& \(8{.}551\)				& \(4{.}2\) & \(22{.}1\) \\
  &	(No.\,149)			& 							& 					& 							&			&			\\
2 & Journal of Physical Che-  & \(0{.}370\)			& \(0{.}313\) 		& \(6{.}187\) 				& \(12{.}5\) & \(46{.}2\) \\
  &	mistry C (No.\,109)			& 					& 					&							&			&			\\
3 & Physical Review E 			& \(0{.}348\) & \(0{.}069\) & \(7{.}307\)	& \(36{.}3\) & \(23{.}4\) \\
  &	(No.\,184)					&			  & 			& 				&			 &			\\
4 & IEEE Transactions on  		& \(0{.}308\) & \(0{.}043\) & \(7{.}760\)	&\(47{.}1\) & \(8{.}4\) \\
  &	Magnetics (No.\,73)			& 			  & 			&				&			&			\\
\hline
\end{tabular}
\end{table*}

\begin{figure}[ht!]
\centering
  \includegraphics[width=6.5cm]{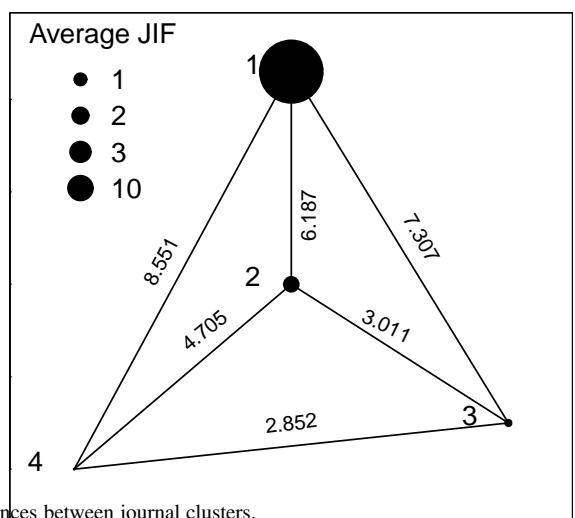}
\caption{Hausdorff distances between journal clusters.}
\label{fig:2}     
\end{figure}

Calculated values for Hausdorff distances Eq.~\eqref{hausdorff} between the clusters are shown in Fig.~\ref{fig:2}. The distance between cluster~1 and 4 is maximal. Clusters~3 and 4 are the closest. In what follows, we shall see that these distances are not accidental. They are consistent with some other properties of physical journals.

\begin{figure}[ht!]
\centering
  \includegraphics[width=\linewidth]{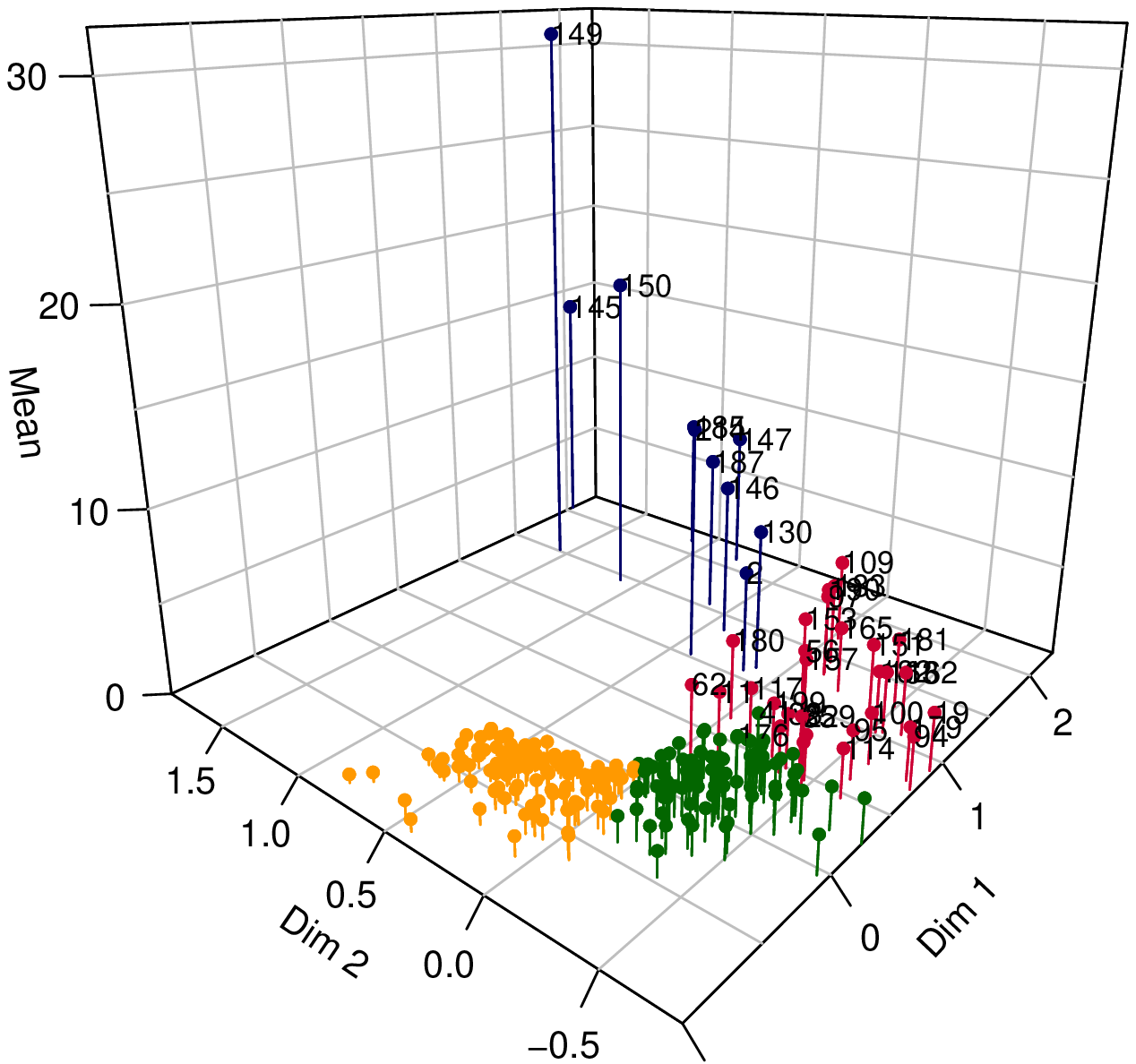}
\caption {MDS ordination and mean of citations of physics journals. MDS created with the use of the PROXSCAL algorithm \cite{Borg_2013} and distances \(\rho (\xi, \eta)\) (S\,--\,Stress \(= 0{.}224\)). Journals having mean of citations higher than \(2{.}5\) marked by numbers. List of journals see in Tab.~\ref{tab1}.}
\label{fig:3}      
\end{figure}

\begin{figure}[ht!]
\centering
  \includegraphics[width=\linewidth]{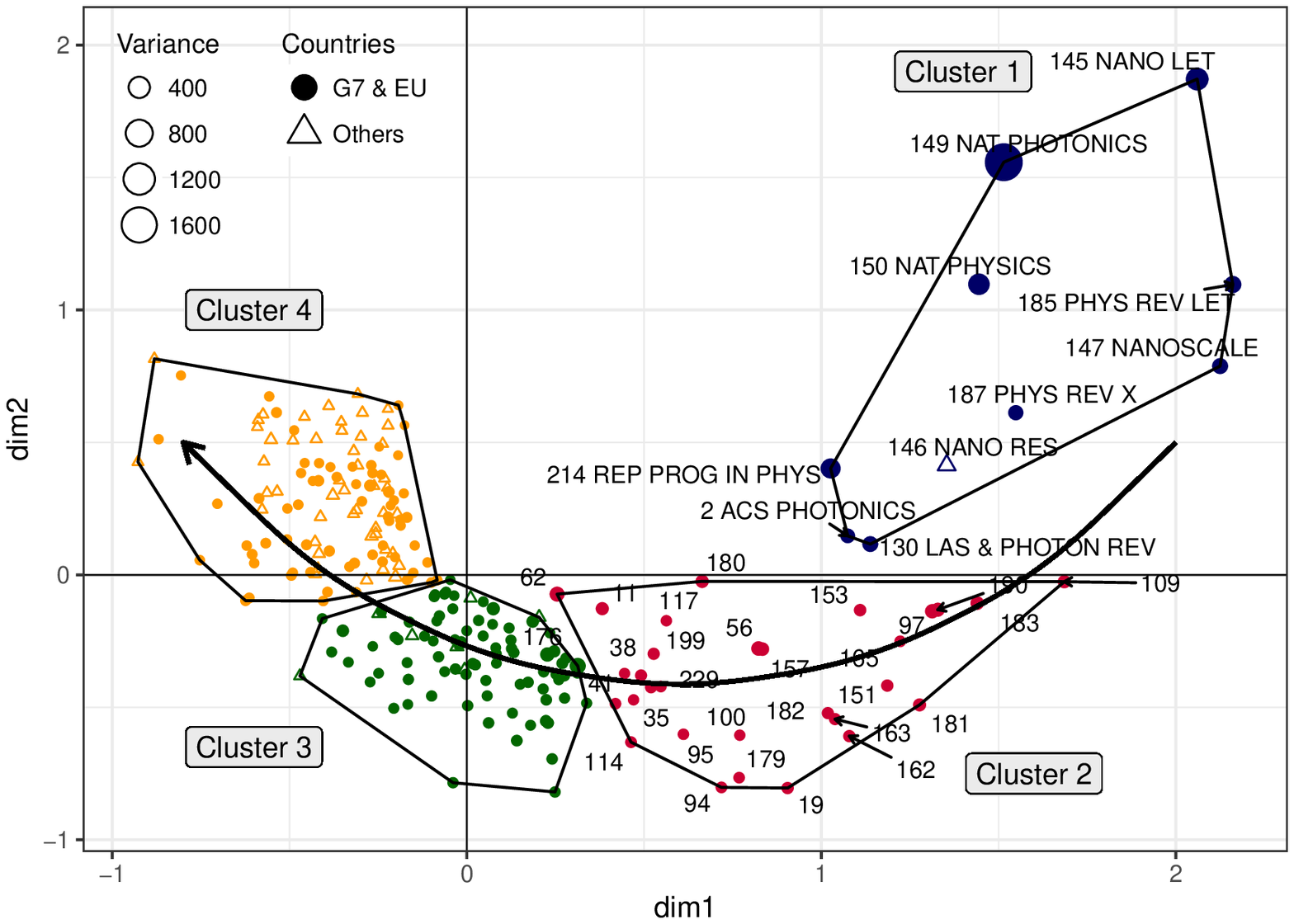}
\caption {MDS plot of physics journals. Journals having mean of citations higher than \(2{.}5\) marked by numbers. Arrow indicates direction of JIF decreasing. List of journals see in Tab.~\ref{tab1}.}
\label{fig:4}      
\end{figure}

The visualization of the SPJ is based on the use of multidimensional scaling (or MDS) (cf. \cite{Zhu:2015:TAI:2811411.2811475, Hierarchical_networks, Leydesdorff2017}). MDS is a technique that represents similarity data as distances between points in low-dimensional geometric space \cite{Borg_2013}. It makes possible graphically display relations among objects to analyze and visualize the pattern of similarity between them. It enables the researcher visually explore the structure of data. We applied MDS method to the distance matrix \(\bigl( \rho (\xi, \eta)\bigr)_{\xi, \eta =1}^{\xi , \eta = 240}\) to construct a two-dimensional common space of physics journals, which satisfactorily describes the initial matrix of distances between journals. This reduces a system of interconnected differences between journals to two generalized ``axes'', shown in Fig.~\ref{fig:3} and Fig.~\ref{fig:4}.

Journals located in the different clusters vary by size, citation level, JIF, national affiliations, and so on. As a result, they enjoy different levels of popularity and prestige (cf. \cite{Ferrer-Sapena2016}). In fact, when moving clockwise from cluster~1 to cluster~4 the properties of the journals changes monotonically. The average value of JIF for journals of cluster~1 is equal to \(12{.}256\), for journals of cluster~2~--- \(3{.}404\); for journals of cluster~3 and cluster~4 the average value of JIF is \(1.843\) and \(0.878\) respectively. The total number of citations received by a journal in the WOS CC changes analogously. The average value for the indicator total cites for journals of cluster~1 is equal to \(65{,}186\); for publications of cluster~2~--- \(56{,}589\). The lowest average value of total cites is observed among the journals of cluster~3 and cluster~4, where the value of the indicator is \(12{,}132\) and \(2909\) respectively.

The highest average value of variance can be observed in cluster~1 where it is \(308{.}3\). The differentiation of physics journals by level of citation in this cluster is very high. In this respect it differs from other clusters, which are more homogeneous. For the rest, the average variance in the level of citation is \(21{.}1\), \(8{.}2\) and \(2{.}3\) respectively.

The indicator for the number of citable items in two years behaves somewhat differently. On average, the journals of cluster~1 publish \(1342\) papers, and the journals of cluster~2~--- \(2535\). The figures for journals of cluster~3 and cluster~4 are significantly less; on average, they publish \(844\) and \(508\) citable items in two years.

We can note an important feature of the journals in cluster~1. By comparison with the others, they are relatively new. The average year of their foundation is 1995. The average year of the foundation of journals of cluster~2 is 1971, cluster~3 and cluster~4 is 1978 and 1979 respectively. The journals of cluster~1 are mainly products of the era of the internet and bibliometric databases. They are more in line with modern tools for the dissemination of scientific information than other journals. This may be one of the factors that contributes to their high bibliometric indicators. The journals in cluster~1 are ``journals of quality'', if one goes by their citation statistics.

The SPJ acts as a kind of meta-language for the description of journals, which enables us to identify regularities through analysis of their properties in a fixed ``alphabet'' of clusters. Thus, when it comes to publishers of journals, certain regularities can be observed when moving in a clockwise direction through the common space of physics journals.

Although publishers' sophisticated strategies are constantly changing the landscape, we attempt to map these regularities. Cluster~1 is dominated by journals which belong to academic societies: American Chemical Society, Institute of Physics, Royal Society of Chemistry, and Tsinhua University. The publishers Springer Nature and Wiley\,--\,Blackwell are the exception to this. As we move from cluster~1 to cluster~4, the influence of scientific publishing companies increases. 

Elsevier, Springer Nature and Wiley\,--\,Blackwell have a share of \(36{.}7\%\) of all journals in the cluster~2. However, leading physics societies also enjoy a significant position. Journals published by Institute of Physics account for \(23{.}3\%\), while American Physical Society and American Institute of Physics have shares of \(16{.}8\%\) and \(6{.}7\%\) respectively. Clusters~2 and 3 can be considered as the ``world of Elsevier''. The large majority of this publisher's journals (\(77{.}3\%\)) can be found in these clusters. In cluster~3, the share of journals owned by large multinational publishers increases, while the share of journals belonging to professional societies decreases. The share of journals published by Institute of Physics, American Physical Society and American Institute of Physics is just \(19{.}1\%\) in cluster~3, while the proportion of journals belonging to Elsevier, Springer Nature and Wiley\,--\,Blackwell is \(51{.}7\%\).

Cluster~4 differs substantially from the three preceding clusters. The first striking characteristic is the concentrated presence of the scientific publishing company Maik Nauka, which publishes translations of Russian journals. All of this publisher's journals can be found in cluster~4, accounting for \(14{.}2\%\) of all journals in this cluster. Almost all journals published by World Scientific can also be found here~--- \(8{.}0\%\) of the total. The share of Springer Nature is \(9{.}7\%\) and Elsevier is \(8{.}9\%\).

Due to the strong internationalization of the publishing process, it is not possible to straightforwardly classify physics journals in terms of national affiliation. However, it is possible to trace the distinction between ``global'' and ``local'', which is formed on the principle of ``the West and the rest''. In moving from cluster~1 to cluster~4 the number of countries to which a journal can be ascribed rises. In cluster~1 the only countries represented are the USA, Great Britain, Germany and China. Cluster~2 contains five countries, of which the Netherlands (to which Elsevier is affiliated) occupies a prominent place. The most international cluster in terms of publisher affiliation is 4, which contains journals affiliated with \(23\) countries. Besides the USA and Great Britain, Russia and Singapore play a significant role in this cluster (home respectively to the publishers Maik Nauka and World Scientific).

The SPJ can be treated, amongst other things, as a projection of competitive relations within the triad ``transnational publishers~--- professional physics societies~--- regional publishers'' (cf. \cite{10.1371/journal.pone.0127502, Greshake_2017}):
\begin{itemize}
\item In cluster~1 professional physics societies are dominant.
\item In cluster~2 there is interference of professional societies and transnational publishers.
\item Cluster~3 is dominated by transnational publishers, while professional physics societies lose their dominance, and have a relatively weak representation.
\item In cluster~4 the dominant players are regional publishers from Asia, Latin American and the former Soviet Union, while the role of professional physics societies and transnational publishers is insignificant. The journals of the fourth cluster are those which stem from ``non-Western'' science, which are less well adapted than others to the demands of global physics and the WoS CC.
\end{itemize}

\begin{figure}[ht!]
\centering
  \includegraphics[width=12.9cm]{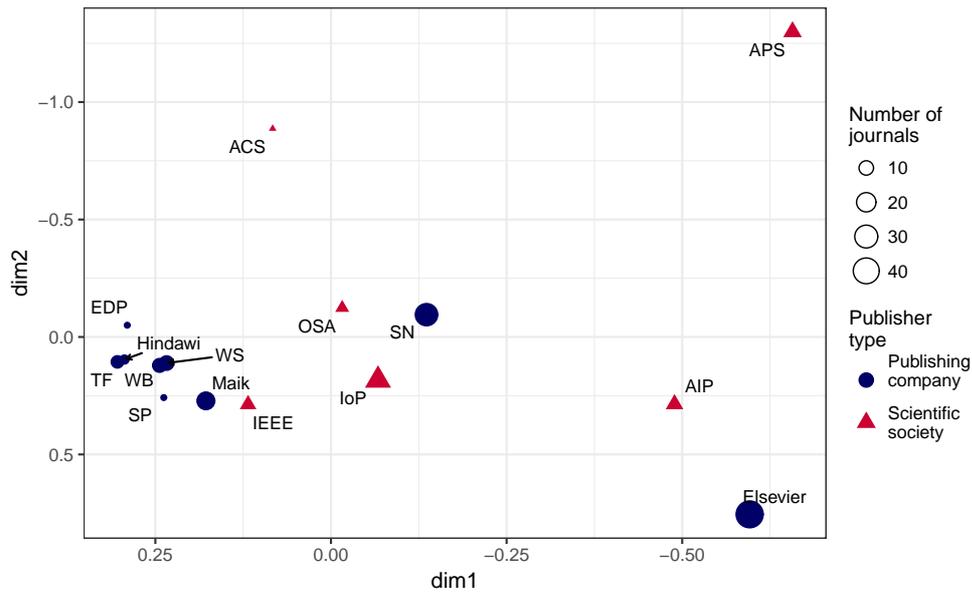}
\caption {MDS plot of 15 scientific publishers. Plot of the first and second coordinates of multidimensional scaling analysis (with the use of the PROXSCAL algorithm \cite{Borg_2013} of Hausdorff distances between publishers (S\,--\,Stress \(= 0{.}509\)). Publishers: ACS (American Chemical Society), AIP (American Institute of Physics), APS (American Physical Society), EDP (EDP Sciences), Elsevier (Elsevier), Hindawi (Hindawi Publishing), IEEE (Institute of Electrical and Electronics Engineers ), IoP (Institute of Physics), Maik (Maik Nauka/Interperiodica Publishing), OSA (Optical Society of America), SN (Springer Nature), SP (Science Press), TF (Taylor \& Francis), WB (Wiley-Blackwell), WS (World Scientific).}
\label{fig:5}      
\end{figure}

While long-term strategies of merger and acquisition can disrupt patterns of similarity between publishers, the citation rates of journals belonging to one publisher often show statistical similarities. We analyzed \(15\) main publishers from the sample, which accounted for \(73{.}3\%\) of all journals. In order to represent relationships between the scientific publishers, the MDS method was applied. Fig.~\ref{fig:5} is the plot of the first and second coordinates of MDS of the matrix \(\bigl( \rho_{\mathrm{H}}(A, B)\bigr)_{A, B =1}^{A, B = 15}\) of Hausdorff distances between all pairs of the publishers.

\begin{table*}[!htb]
\centering
\caption{Bibliometric characteristics of major publishers}
\label{tab:2}
\begin{tabular}{lllll}
\hline
Publisher & Citations per & Number of			& Publications	in		& Citations \\
		  & publication & journals (\(\%\)) & 2013\,--\,2014 (\(\%\)) & in 2015 (\(\%\)) \\
\hline
ACS & 6.7 & 1.3 & 4.0 & 10.4 \\
APS & 4.2 & 3.8 & 16.3 & 26.4 \\
Springer Nature & 2.9 & 11.7 & 7.8 & 8.9 \\
OSA & 2.8 & 2.1 & 6.4 & 7.0 \\
AIP & 2.4 & 3.3 & 13.5 & 12.7 \\
IoP & 2.3 & 8.3 & 6.7 & 5.9 \\
Wiley\,--\,Blackwell & 2.0 & 3.8 & 1.8 & 1.4 \\
Elsevier & 1.9 & 18.3 & 16.1 & 11.8 \\
EDP Sciences & 1.7 & 1.3 & 1.2 & 0.7 \\
IEEE & 1.3 & 2.9 & 3.8 & 1.9 \\
Taylor \& Francis & 1.0 & 2.9 & 1.2 & 0.5 \\
World Scientific & 0.9 & 4.2 & 1.6 & 0.6 \\
Maik Nauka & 0.7 & 6.7 & 2.8 & 0.7 \\
Hindawi & 0.7 & 1.7 & 0.6 & 0.2 \\
Science Press & 0.6 & 1.3 & 1.7 & 0.4 \\
Others & 1.9 & 26.7 & 14.6 & 10.6\\
\hline
\end{tabular}
\end{table*}

The \(\mathrm{dim1}\) axis in Fig.~\ref{fig:5} opposes global publishers to local ones. On the right pole are Elsevier, the American Physical Society and the American Institute of Physics. On the left pole are all regional publishers EDP Press (France), Hindawi (Egypt), Science Press (China), World Scientific, Maik Nauka (Russia) and some global ones: Wiley\,--\,Blackwell and Taylor \& Francis.

The \(\mathrm{dim2}\) axis in Fig.~\ref{fig:5} provides a visualization of the differences between high and low cited publishers. At the top of the dim2 axis are the American Physical Society and the American Chemical Society having \(6{.}7\) and \(4{.}2\) citations per publication respectively. At the bottom of the \(\mathrm{dim2}\) axis are Elsevier (\(1{.}9\) citations per publication) and the Institute of Electrical and Electronics Engineers (\(1{.}3\) citations per publication). 

It can be suggested from Fig.~\ref{fig:5} that the space of physics publishers is structured around two fundamental oppositions:
\begin{enumerate}
\item The first opposition centres around global publishers on the one hand, contrasted with mainly regional publishers on the other hand.
\item The dominant role in the second key opposition is played by the differences separating publishers with highly cited journals and publishers with low cites journals.
\end{enumerate}

\section*{Conclusions}
In order not to lose its \emph{raison d'\^{e}tre}, scientometrics must fulfil the main cognitive task of any form of science~--- explanation. If scientometrics is to have a claim to scientific authority, the task of any potential scientometric concept must be to explain the facts in terms of its own theory \cite{hj2016}. Scientific explanation establishes a logical connection between an individual scientometric fact and the totality of relevant facts, and includes general knowledge in the composition of an individual fact. In scientometrics, ``explaining physics journals'' means establishing the connections and relationships that determine the essential features of these journals. In this paper, the set of relations of physics journals is fixed in the form of a topological structure on the sample \(S\). In our case, explaining physics journals means including the corresponding sample of journals in the topological structure \(\mathcal{T}\) of given connections, relationships and dependencies (cf. \cite{PhysRevE.95.012324}).

The system of four clusters \(\bigl(\mathrm{Cl}_{j}\bigr)_{j = 1}^{j = 4}\), i.e. \(\mathcal{T}\) on \(S\), serves as the conditions for and means of explanation and it is within the framework of this structure that the properties of physics journals (and the differences between them) are formulated and described. In this paper, the system of indiscrete clusters of physics journals is a natural generalization of the concept of closeness of a journal to a totality of journals. Furthermore, the deviation \(d(\xi)\) is a formalization of the distance from the journal \(\xi\) to the sample \(S\). In this regard, if JIF is expressed in terms of \(d(\xi)\), one clearly sees a connection between JIF and \(\mathcal{T}\). As such, the proposed explanation of JIF makes available one topological basis for the linkage of the quantities \(\mathrm{JIF}(\xi)\) and \(d(\xi)\). On the other hand, the deviation \(d(\xi)\) seems to be of significance in the structure of SPJ, while being intimately related to the minimal neighborhood base of \((S, \rho)\).

Following our findings, we can imagine that \(\mathcal{T}\) on \(S\) agrees with the ranking of physics journals according to their JIF (cf. \cite{Huang2017173}), because \(S\) is partially ordered by \(\precsim\) \eqref{po}. The deeper significance of JIF will become apparent when proposed methodology gives a framework for thinking of JIF as a result of composite structure of SPJ, partly ordered and partly topological, and the two parts are related to each other by natural rules.

The suggested approach provides a topology structure for the bibliometric grouping of physics journals; in this structure continuous passing from any EDF to the others EDFs makes sense. The SPJ is not only a geometrical locus of relations between citation distributions, but also between journals and publishers as social forces which aim at transforming these relations. If certain underlying causes are responsible for certain differences between the EDFs, then the structure corresponding to these underlying causes must be found in the field of physics. If clusters of physics journals show certain differences from each other, then such differences are typical for the underlying causes that are responsible for the existence of these clusters.

Journals with high JIFs vary greatly, and are located at a considerable distance from each other in space (see Fig.~\ref{fig:4}). In contrast, journals with low JIF values are located close to each other. Here we can note a regularity: EDFs better characterize ``prestigious'' physics journals as opposed to their ``less prestigious'' counterparts (cf. \cite{Stern:2013, Mila_2017}). The relationships of these ``less prestigious'' journals are best described by the indicators ``publisher'' and ``national affiliation''. More generally, the meaning of EDFs for journals varies depending on their source cluster. SPJ is non-isotropic: not only do physics journals' bibliometric properties change depending on their location in this space, the social meaning of those properties also alters.

An essential characteristic of a good explanation of JIF is that it tells a story about a totality of scientific journals. This paper has emphasized, that it would seem reasonable to consider a scientific journal as a representative of the family of journals, all coexisting interdependently in citation field. The topological structure sheds light not only on physics journals as direct forms of scientific communication, but also reveals the structuring role of publishers, something which is not usually highlighted. Analysis of the topological structure allows us to advance the hypothesis that the positions of physics journals are determined, among others, by a system of relations between publishers.

It is difficult to conceive of an operationalization of SPJ free of scientometric critic. It is obvious that SPJ expresses a complex phenomenon. However, whilst setting aside this complexity, in scientometrics it is possible to work with an individual aspect of SPJ. The multifaceted nature of SPJ indicates the relatively complex structure of the concept. The differences between the elements of this structure can be so great that this in itself can justify the study of the individual aspects as isolated scientometric constructs. The topological structure is one of these constructs. We hope that this paper offers a new and valid way of studying scientific journals to anybody who treats it mathematically.
\section*{Acknowledgments}
The article was prepared within the framework of the Basic Research Program at the National Research University Higher School of Economics (HSE) and supported within the framework of a subsidy by the Russian Academic Excellence Project ``5-100''.

\bibliographystyle{spbasic}      

\section*{Appendix~1: Physics journals in the sample}
See Table~\ref{tab1}

\begin{table}[!htb]
\caption{Physics journals in the sample}%
\label{tab1}
\scriptsize
\begin{tabular}{lllllllll}
\hline
No. & Journal title & Cl & No. & Journal title & Cl & No. & Journal title & Cl \\ 
\hline
1 & ACOUST PHYS+ & 4 & 81 & INT J MOD PHYS A & 3 & 161 & OPT COMMUN & 3 \\
2 & ACS PHOTONICS & 1 & 82 & INT J MOD PHYS B & 4 & 162 & OPT EXPRESS & 2 \\
3 & ACTA ACUST UNITED AC & 4 & 83 & INT J MOD PHYS C & 4 & 163 & OPT LETT & 2 \\
4 & ACTA PHYS POL A & 4 & 84 & INT J MOD PHYS E & 4 & 164 & OPTIK & 4 \\
5 & ACTA PHYS POL B & 4 & 85 & INT J PHOTOENERGY & 4 & 165 & ORG ELECTRON & 2 \\
6 & ACTA PHYS SIN-CH ED & 4 & 86 & INT J QUANTUM INF & 4 & 166 & PHASE TRANSIT & 4 \\
7 & ADV COND MATTER PHYS & 4 & 87 & INT J THEOR PHYS & 4 & 167 & PHIL MAG LETT & 4 \\
8 & ADV HIGH ENERGY PHYS & 4 & 88 & INVERSE PROBL & 3 & 168 & PHILOS T R SOC A & 3 \\
9 & ADV MATH PHYS & 4 & 89 & IONICS & 3 & 169 & PHOTONIC NANOSTRUCT & 3 \\
10 & AM J PHYS & 4 & 90 & JPN J APPL PHYS & 4 & 170 & PHYSICA A & 3 \\
11 & ANN PHYS-BERLIN & 2 & 91 & JETP LETT+ & 4 & 171 & PHYSICA B & 3 \\
12 & ANN HENRI POINCARE & 3 & 92 & J APPL MECH TECH PH+ & 4 & 172 & PHYSICA C & 4 \\
13 & ANN PHYS-NEW YORK & 3 & 93 & J APPL PHYS & 3 & 173 & PHYSICA D & 3 \\
14 & APPL ACOUST & 3 & 94 & J CHEM PHYS & 2 & 174 & PHYSICA E & 3 \\
15 & APPL OPTICS & 3 & 95 & J COMPUT PHYS & 2 & 175 & PHYS SCRIPTA & 4 \\
16 & APPL PHYS A-MATER & 3 & 96 & J EXP THEOR PHYS+ & 4 & 176 & PHYS STATUS SOLIDI-R & 3 \\
17 & APPL PHYS B-LASERS O & 3 & 97 & J HIGH ENERGY PHYS & 2 & 177 & PHYS STATUS SOLIDI A & 3 \\
18 & APPL PHYS EXPRESS & 3 & 98 & J INFRARED MILLIM W & 4 & 178 & PHYS STATUS SOLIDI B & 3 \\
19 & APPL PHYS LETT & 2 & 99 & J LOW TEMP PHYS & 4 & 179 & PHYS REV A & 2 \\
20 & APPL RADIAT ISOTOPES & 4 & 100 & J LUMIN & 2 & 180 & PHYS REV APPL & 2 \\
21 & ARCH ACOUST & 4 & 101 & J MAGN & 4 & 181 & PHYS REV B & 2 \\
22 & BRAZ J PHYS & 4 & 102 & J MAGN MAGN MATER & 3 & 182 & PHYS REV C & 2 \\
23 & B LEBEDEV PHYS INST+ & 4 & 103 & J MATH PHYS & 4 & 183 & PHYS REV D & 2 \\
24 & CAN J PHYS & 4 & 104 & J MOD OPTIC & 4 & 184 & PHYS REV E & 3 \\
25 & CENT EUR J PHYS & 4 & 105 & J NANOPHOTONICS & 3 & 185 & PHYS REV LETT & 1 \\
26 & CHALCOGENIDE LETT & 4 & 106 & J NON-CRYST SOLIDS & 3 & 186 & PHYS REV SPEC TOP-AC & 3 \\
27 & CHAOS & 3 & 107 & J OPTICS-UK & 3 & 187 & PHYS REV X & 1 \\
28 & CHAOS SOLITON FRACT & 3 & 108 & J PHASE EQUILIB DIFF & 4 & 188 & PHYS-USP+ & 3 \\
29 & CHINESE J CHEM PHYS & 4 & 109 & J PHYS CHEM C & 2 & 189 & PHYS LETT A & 3 \\
30 & CHINESE J PHYS & 4 & 110 & J PHYS-CONDENS MAT & 3 & 190 & PHYS LETT B & 2 \\
31 & CHIN OPT LETT & 3 & 111 & J PHYS A-MATH THEOR & 3 & 191 & PHYS ATOM NUCL+ & 4 \\
32 & CHINESE PHYS B & 3 & 112 & J PHYS CHEM SOLIDS & 3 & 192 & PHYS FLUIDS & 3 \\
33 & CHINESE PHYS C & 4 & 113 & J PHYS B-AT MOL OPT & 3 & 193 & PHYS PART NUCLEI+ & 4 \\
34 & CHINESE PHYS LETT & 4 & 114 & J PHYS D APPL PHYS & 2 & 194 & PHYS PLASMAS & 3 \\
35 & CLASSICAL QUANT GRAV & 2 & 115 & J PHYS G NUCL PARTIC & 3 & 195 & PHYS SOLID STATE+ & 4 \\
36 & COMMUN COMPUT PHYS & 4 & 116 & J PLASMA PHYS & 4 & 196 & PLASMA PHYS CONTR F & 2 \\
37 & COMMUN MATH PHYS & 3 & 117 & J RHEOL & 2 & 197 & PLASMA PHYS REP+ & 4 \\
38 & COMMUN NONLINEAR SCI & 2 & 118 & J STAT MECH-THEORY E & 3 & 198 & PLASMA SCI TECHNOL & 4 \\
39 & COMMUN THEOR PHYS & 4 & 119 & J STAT PHYS & 3 & 199 & PLASMA SOURCES SCI T & 2 \\
40 & CR PHYS & 3 & 120 & J SUPERCOND NOV MAGN & 4 & 200 & POWDER DIFFR & 4 \\
41 & COMPUT PHYS COMMUN & 3 & 121 & J SYNCHROTRON RADIAT & 3 & 201 & PRAMANA-J PHYS & 4 \\
42 & CONDENS MATTER PHYS & 4 & 122 & J ACOUST SOC AM & 3 & 202 & P ROY SOC A-MATH PHY & 3 \\
43 & CONTRIB PLASM PHYS & 3 & 123 & J EUR OPT SOC-RAPID & 4 & 203 & PROG NAT SCI-MATER & 3 \\
44 & CRYOGENICS & 4 & 124 & J KOREAN PHYS SOC & 4 & 204 & PROG THEOR EXP PHYS & 4 \\
45 & CURR APPL PHYS & 3 & 125 & J OPT SOC AM A & 3 & 205 & QUANTUM ELECTRON+ & 4 \\
46 & DOKL PHYS & 4 & 126 & J OPT SOC AM B & 3 & 206 & QUANTUM INF COMPUT & 3 \\
47 & ECS J SOLID STATE SC & 3 & 127 & J OPT SOC KOREA & 4 & 207 & QUANTUM INF PROCESS & 3 \\
48 & ECS SOLID STATE LETT & 4 & 128 & J PHYS SOC JPN & 3 & 208 & RADIAT EFF DEFECT S & 4 \\
49 & ENTROPY-SWITZ & 3 & 129 & J X-RAY SCI TECHNOL & 4 & 209 & RADIAT MEAS & 4 \\
50 & EPL-EUROPHYS LETT & 3 & 130 & LASER PHOTONICS REV & 1 & 210 & RADIAT PHYS CHEM & 3 \\
51 & EUR J PHYS & 4 & 131 & LASER PART BEAMS & 3 & 211 & RADIOPHYS QUANT EL+ & 4 \\
52 & EUR PHYS J-APPL PHYS & 4 & 132 & LASER PHYS & 4 & 212 & REND LINCEI-SCI FIS & 4 \\
53 & EUR PHYS J-SPEC TOP & 3 & 133 & LASER PHYS LETT & 3 & 213 & REP MATH PHYS & 4 \\
54 & EUR PHYS J A & 3 & 134 & LETT MATH PHYS & 3 & 214 & REP PROG PHYS & 1 \\
55 & EUR PHYS J B & 4 & 135 & LOW TEMP PHYS+ & 4 & 215 & REV MEX FIS & 4 \\
56 & EUR PHYS J C & 2 & 136 & MAGNETOHYDRODYNAMICS & 4 & 216 & RHEOL ACTA & 3 \\
57 & EUR PHYS J D & 4 & 137 & MECH SOLIDS+ & 4 & 217 & ROM J PHYS & 4 \\
58 & EUR PHYS J E & 3 & 138 & METROL MEAS SYST & 3 & 218 & ROM REP PHYS & 4 \\
59 & EUR PHYS J PLUS & 4 & 139 & MICRO NANO LETT & 4 & 219 & RUSS J PHYS CHEM B+ & 4 \\
60 & FERROELECTRICS & 4 & 140 & MOD PHYS LETT A & 4 & 220 & RUSS PHYS J+ & 4 \\
61 & FEW-BODY SYST & 4 & 141 & MOD PHYS LETT B & 4 & 221 & SCI CHINA PHYS MECH & 3 \\
62 & FORTSCHR PHYS & 2 & 142 & MOL PHYS & 3 & 222 & SEMICOND SCI TECH & 3 \\
63 & FOUND PHYS & 4 & 143 & MOSC U PHYS B+ & 4 & 223 & SEMICONDUCTORS+ & 4 \\
64 & FRONT PHYS-BEIJING & 3 & 144 & NANO & 4 & 224 & SOLID STATE ELECTRON & 3 \\
65 & GEN RELAT GRAVIT & 3 & 145 & NANO LETT & 1 & 225 & SOLID STATE COMMUN & 3 \\
66 & HIGH ENERG DENS PHYS & 3 & 146 & NANO RES & 1 & 226 & SOLID STATE IONICS & 2 \\
67 & HIGH PRESSURE RES & 4 & 147 & NANOSCALE & 1 & 227 & SOLID STATE SCI & 3 \\
68 & HIGH TEMP+ & 4 & 148 & NANOSCALE RES LETT & 3 & 228 & STUD HIST PHILOS M P & 4 \\
69 & IEEE J QUANTUM ELECT & 3 & 149 & NAT PHOTONICS & 1 & 229 & SUPERCOND SCI TECH & 2 \\
70 & IEEE PHOTONICS J & 3 & 150 & NAT PHYS & 1 & 230 & SUPERLATTICE MICROST & 3 \\
71 & IEEE PHOTONIC TECH L & 3 & 151 & NEW J PHYS & 2 & 231 & SURF INTERFACE ANAL & 4 \\
72 & IEEE T APPL SUPERCON & 4 & 152 & NUCL DATA SHEETS & 4 & 232 & SURF REV LETT & 4 \\
73 & IEEE T MAGN & 4 & 153 & NUCL FUSION & 2 & 233 & SURF SCI & 3 \\
74 & IEEE T NUCL SCI & 3 & 154 & NUCL INSTRUM METH A & 4 & 234 & SYMMETRY INTEGR GEOM & 4 \\
75 & IEEE T PLASMA SCI & 4 & 155 & NUCL INSTRUM METH B & 3 & 235 & SYNTHETIC MET & 2 \\
76 & IMAGING SCI J & 4 & 156 & NUCL PHYS A & 4 & 236 & TECH PHYS+ & 4 \\
77 & INDIAN J PHYS & 4 & 157 & NUCL PHYS B & 2 & 237 & TECH PHYS LETT+ & 4 \\
78 & INDIAN J PURE AP PHY & 4 & 158 & NUCL SCI TECH & 4 & 238 & THEOR MATH PHYS+ & 4 \\
79 & INFRARED PHYS TECHN & 3 & 159 & NUKLEONIKA & 4 & 239 & THERMOPHYS AEROMECH+ & 4 \\
80 & INT J GEOM METHODS M & 4 & 160 & OPT REV & 4 & 240 & WAVE MOTION & 3 \\
\hline
\end{tabular}
\end{table}
\normalsize

\end{document}